\documentclass[journal=jpcafk,manuscript=article]{achemso}
\usepackage[version=3]{mhchem}
\setkeys{acs}{articletitle = true}
\usepackage{titlecaps}
\Addlcwords{the of into}
\usepackage[T1]{fontenc} 
\usepackage{enumitem}
\usepackage{graphicx}

\newcommand*{\citen}[1]{%
  \begingroup
    \romannumeral-`\x 
    \setcitestyle{numbers}%
    \cite{#1}%
  \endgroup   
}

\author{Debasish Koner}\affiliation[University of Basel]
       {Department of Chemistry, University of Basel,
         Klingelbergstrasse 80, 4056 Basel, Switzerland}
\author{Raymond J. Bemish} \affiliation{Air Force Research Laboratory,
  Space Vehicles Directorate, Kirtland AFB, New Mexico 87117, USA}
\author{Markus
  Meuwly}\email{m.meuwly@unibas.ch}\affiliation[University of Basel]
       {Department of Chemistry, University of Basel,
         Klingelbergstrasse 80, 4056 Basel, Switzerland}

\date{\today}

\title[]{Dynamics on Multiple Potential Energy Surfaces: Quantitative
  Studies of Elementary Processes Relevant to Hypersonics}

\abbreviations{}

\begin{document}

\begin{abstract}
The determination of thermal and vibrational relaxation rates of
triatomic systems suitable for application in hypersonic model
calculations is discussed. For this, potential energy surfaces for
ground and electronically excited state species need to be computed
and represented with high accuracy and quasiclassical or quantum
nuclear dynamics simulations provide the basis for determining the
relevant rates. These include thermal reaction rates, state-to-state
cross sections, or vibrational relaxation rates. For exemplary systems
- [NNO], [NOO], and [CNO] - all individual steps are described and a
literature overview for them is provided. Finally, as some of these
quantities involve considerable computational expense, for the example
of state-to-state cross sections the construction of an efficient
model based on neural networks is discussed. All such data is required
and being used in more coarse-grained computational fluid dynamics
simulations.
\end{abstract}

\section{Introduction}
In hypersonic flight an object traveling at high speed through an
atmosphere will dissipate large amounts of energy to the surrounding
gas and generate highly non-equilibrium conditions with respect to
occupation of translational, rotational, vibrational, and electronic
degrees of freedom of the surrounding molecules. Typically, the
energies (and hence temperatures) are sufficiently high to dissociate
small molecules such as N$_2$ and O$_2$. At these extremes, the energy
dissipated due to chemistry can be comparable to shock and skin
friction interactions. For Earth's atmosphere the main constituents of
the air at altitudes for which the medium is sufficiently dense for
frequent collisions (30 km to 60 km above sea level, i.e. troposphere
and stratosphere) are O$_2$, N$_2$, and NO. Mars, Titan, Venus and
other planets with dense atmospheres have significantly more complex
polyatomic species to consider.\\

\noindent
Hypersonic flight is an endeavor of grand scale.  A hypersonic vehicle
covers speeds of kilometers per second and experiences surface
temperatures only limited by the vaporization temperature of its outer
shell, is exposed to tens of MW/m$^2$ of heating and generates a bow
shock with temperatures in excess of 20000 K. In subsonic flight, the
dynamics is driven by the flow across a surface. At supersonic speeds,
the dissipation of the flow is dominated by the generation of shock
waves.  At hypersonic velocities, typically considered as above Mach
5,\cite{cummings:2003} the flow is dominated by chemistry. In the case
of Earth's atmosphere, this is primarily the combustion of
nitrogen. Under such extreme flow conditions, local heating, surface
ablation, control surface authority and plasma formation are directly
sensitive to the energy distribution in molecules and atoms, spanning
a range in time and space of $10^{12}$ between atomic and molecular
collisions and macroscopic changes in the morphology or composition of
matter.\\

\noindent
There have been several reviews and monographs, especially in the
aerospace engineering literature about the historical development of
chemistry models for hypersonic
flow.\cite{sarma:2000,walpot:2012,dsmc:2017} Briefly, the model
development has been driven from top down by the two common approaches
used to solve the flow problem: computational fluid dynamics
(CFD)\cite{walpot:2012} and direct simulation Monte Carlo
(DSMC).\cite{dsmc:2017} Again, it is not the point of this article to
discuss the techniques, there are several others and a multitude of
ways that these two have been implemented to accurately account for
the necessary accommodations that arise from the computational
formulation of the problem. The two approaches however are
fundamentally different and have uses that overlap, but are largely
complementary. The Navier-Stokes equations provide the foundation of
most CFD approaches and have specific requirements, notably the need
for momentum transfer in the fluid and for a differentiable flow
field. This is not however, by definition, met in rarefied flow due to
the low density or in the bow shock due to the discontinuity in the
flow.\\

\noindent
DSMC on the other hand is a probabilistic approach and tracks the
probability of reaction and molecular internal-state outcomes in a
discretized system, without the underlying requirement for viscosity
or differentiability.\cite{dsmc:2017} As an example, a cell within a
grid for DSMC may have collisions and the individual molecules are not
tracked from grid to grid or time step to time step. Rather, they are
tracked as particles with probabilistic outcomes generated at
timesteps and their internal state $(v,j)$ and translational energy
$E_t$ is shared with adjacent voxels. Such an approach is therefore an
excellent choice for high altitude/orbital drag and modeling bow
shock. On the other hand, the computational modeling becomes demanding
at higher particle number density for which the time step and grid
size rapidly decrease to keep the number of events per time step to
the order of 1.\\

\noindent
Both CFD and DSMC require information about the chemistry that occurs
in the flow. For CFD, this is the reaction enthalpy, the reaction
kinetics and the vibrational energy transfer rates.  DSMC typically
uses the reaction cross sections instead of rates as the momentum
transfer in continuum flow drives the system to a Boltzmann
distribution, it is not necessarily the case in a rarefied flow. In
either case, since the vibrational relaxation rate is removed from the
reaction rates, it is possible to have non-equilibrium, where the
vibrational cooling occurs at a significantly different rate than the
translational/rotational or electronic cooling.  In DSMC, since this
can be examined at the state-to-state level, there is also the
possibility of flow solutions that are non-Boltzmann in internal
energy.\\

\begin{figure}[htbp]
\centering
\includegraphics[width=0.99\textwidth]{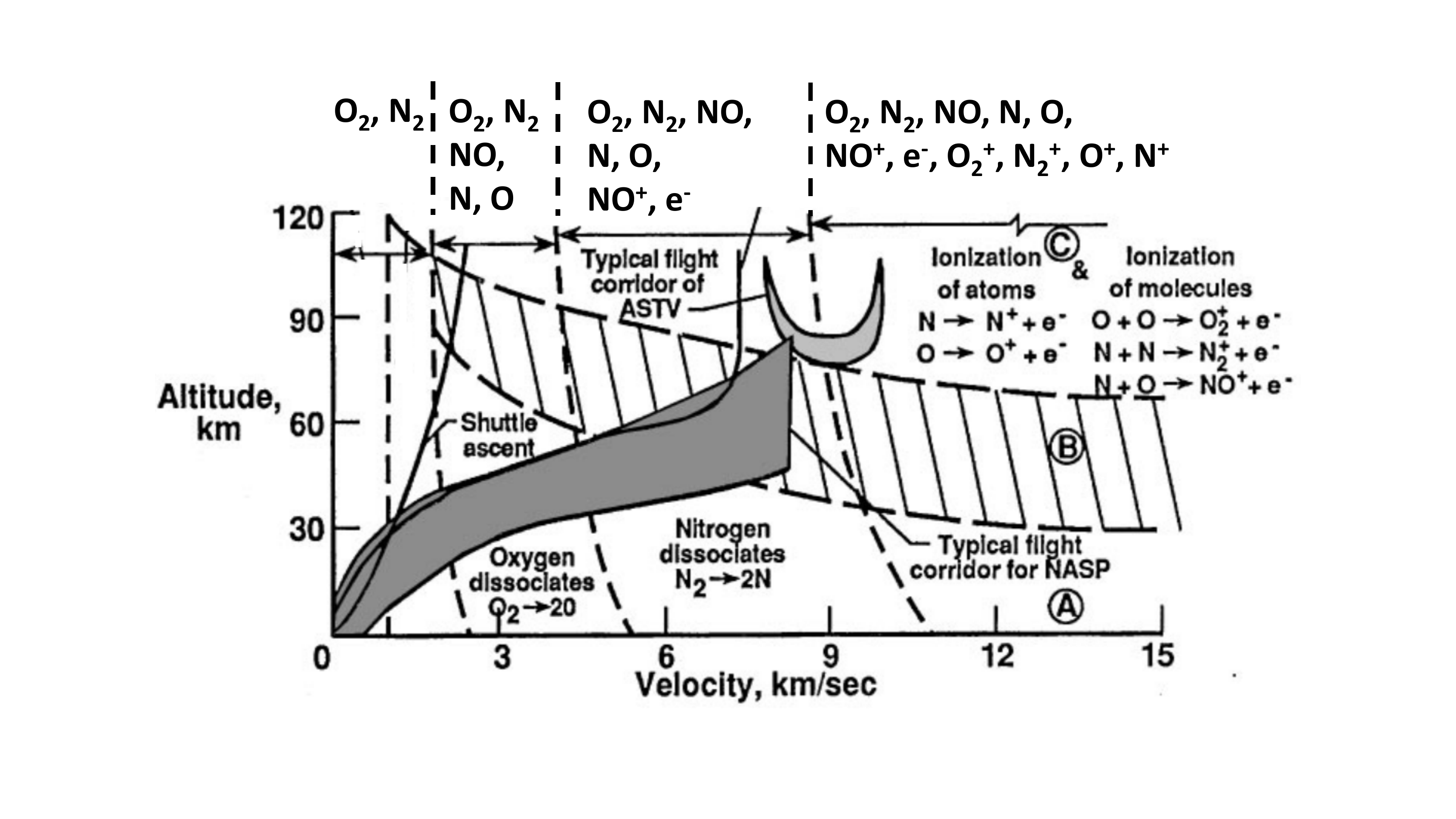}
\caption{Relevant chemical species depending on speed and altitude of
  the traveling vehicle for aeroassisted space transfer vehicle (ASTV)
  and national aero-space plane (NASP), adapted Figure 1 from
  Ref.\cite{sarma:2000}. The top row reports the species involved in
  chemical process and include the so-called, 2-, 5-, 7-, and
  11-species models.\cite{gupta:1990,park:1993} The [NNO] and [NOO]
  species provide a comprehensive model for the non-ionized parts of
  the reaction network. Vibrational relaxation becomes relevant in the
  region of the 5-species model. Region A: chemical and thermal
  equilibrium; region B: chemical non-equilibrium and thermal
  equilibrium; region C: chemical and thermal non-equilibrium.}
\label{fig:chemistry}
\end{figure}

\noindent
The common approach for incorporating chemistry into CFD modeling is
due to Park.\cite{park:1993,park:2001} In a series of publications,
the approach was developed to allow for multiple temperatures, $T_{\rm
  v}$ and $T_{\rm t}$ being the vibrational and
translational/rotational temperatures. In application to kinetics the
temperature is taken to be the geometric mean of these, the so-called
"$T-T_v$ model".  Additionally, following and extending the approach
from Millikan and White,\cite{millikan:1963} an important intuitive
correction established a framework for including the vibrational
relaxation that is required for the generation of vibrational
non-equilibrium.\cite{park:1994} With only minor variations, this
approach has been applied for the last 25 years.  With larger
computational platforms, it has been possible to investigate the
underlying physics on which the Park approach rests. For example, it
was shown for the N$_2$+N and N$_2$ + N$_2$ model that, using the
Millikan-White vibrational relaxation model, the $TT_v$ model predicts
a much faster N$_2$ dissociation for $T \leq 2000$ than that obtained
with direct molecular simulations whereas for $T = 30000$ K the two
models agree.\cite{candler:2016} Additional work on vibrational
relaxation\cite{panesi:2013,MM.arco:2017} shows that there is a clear
difference between the modified Millikan and White model for
vibrational relaxation and what is expected from high fidelity quantum
mechanical or quasiclassical trajectory simulations, by up to 7 orders
of magnitude. Since these rates affect the major chemical species in
the flow, they will at the largest scale even influence the
aerodynamic properties.\\

\noindent
Consequently, accurate state-to-state cross sections are required to
support hypersonic flow modeling. Reliability is a core requirement as
these will influence processes at $10^8$ to $10^{10}$ orders of time
and space higher. This becomes a problem as most of the cross sections
are derived from chemical kinetics, many of which have not
historically been measured at and above 3000K, or can not be measured
at even higher temperatures. Additionally properties like vibrational
relaxation times or the distribution of vibrational and rotational
states often need to be inferred or modelled whereas explicit
determination from rigorous atomistic simulations is likely to provide
less biased quantities that can be used as input for more coarse
grained modeling.\\

\noindent
The present work provides an overview of recent progress, both in
terms of technical improvements and in determining essential
molecular-level information for use in more coarse-grained simulations
and characterization of the systems per-se, for understanding reactive
and non-reactive processes at high temperatures, relevant to the
hypersonic flight regime. The focus is on high-level, extensible data
both, in terms of accuracy and in terms of covering chemical
space. Hence, the methodological ansatz is chosen such that new
reactions can be incorporated seamlessly.\\

\section{Computational Models}
Molecular-level information, such as state-to-state cross sections
$\sigma(\nu,j\rightarrow \nu',j';E_{\rm t})$, thermal rates $k(T)$,
vibrational relaxation (VR) rates, VR times, and the final state
distributions of ro-vibrational states and translational energies that
can be used in more coarse-grained simulations, such as DSMC, can be
obtained by solving the dynamical equations for a number of selected
initial conditions and computing the relevant observables. This is not
only useful for coarse grained models but also a very valuable source
for comparison and interpretation of laboratory-based
experiments. Solving the dynamical equations can be done either by
adopting a quantum mechanical (QM) or a classical mechanical
viewpoint. When using a QM-based method, a time-independent or
time-dependent formalism has to be employed. For classical simulations
the most common approach uses quasi-classical trajectory (QCT)
studies. For both such studies the intermolecular potential energy
surface (PES) encapsulates the totality of physical interactions for a
given electronic state for all atoms involved and provides the most
fundamental level to address the problem at hand. These points are
discussed in some more detail in the next few sections within the
limit that the electronic states are considered separate entities and
that the electrons can move significantly faster than the nuclei
(Born-Oppenheimer approximation).\\

\subsection{Electronic Structure Calculations}
Exploring the electronic structures for different critical
configurations (i.e., stationary points and entrance or exit channels)
is the first and foremost part prior to constructing a
full-dimensional PES. This gives an impression about the number of
electronic states important in studying the collision dynamics of a
particular system and also provides knowledge about the nature of the
electronic wavefunctions at the critical regions. The C-, N-, and
O-containing species show highly multireference character near the
asymptotic regions and single reference methods typically fail to
describe those regions of the PES. Hence, a multi reference
configurations interaction (MRCI) method is necessary to provide an
accurate description of the energetics, in particular for
electronically excited states.\\

\noindent
Complete active space self-consistent field (CASSCF)\cite{wen85:5053,kno85:259,wer80:2342} calculations are
performed prior to MRCI calculations to generate the initial wave
function.  However, single state CASSCF method often fails to converge
near the avoided crossing regions and state averaged (SA) CASSCF
calculations are therefore prescribed.  All the important electronic
states with different possible spin and spatial symmetries are
included in the SA-CASSCF calculations. MRCI calculations are then
performed for a particular state starting from the SA-CASSCF
wavefunction with equal weight on each of the electronic
states. Dynamically weighted SA-CASSCF calculations are also performed
in some cases. Davidson corrections (MRCI+Q)\cite{davidson:1974,wer88:5803,kno88:514}
are used to reduce the
size consistency error. Basis functions are chosen to provide a
healthy balance between accuracy and computational expense. The
augmented Dunning-type correlation consistent polarize triple zeta
(aug-cc-pVTZ)\cite{dun89:1007} are generally enough to give proper description of the
systems. Explicitly correlated MRCI (MRCI-F12) methods can also be
used to reduce errors originating from the finite size of the atomic
basis set.\\

\subsection{Non-reactive and Reactive Potential Energy Surfaces}
Due to continuous improvements of computer architectures and
efficiency gains in the numerical methods, fully-dimensional PESs for
triatomic systems can now be routinely calculated at the
coupled-cluster or multi-reference configuration interaction (MRCI)
level of theory. For smaller electronic systems even full CI (FCI)
treatments with large basis sets are possible.\cite{MM.heh2:2019} A
complete coverage for a triatomic, reactive systems
(A+BC$\rightarrow$AB+C) requires of the order of $10^4$ energy
evaluations. Hence, over the past few years the challenge has
partially shifted away from the computation of reference energies to
representing them.\\

\noindent
Dynamical calculations continuous PESs over all energies accessed by
the simulations. Possibilities to {\it represent} the {\it ab initio}
calculated energies include conventional parametrized fits, the
modified Shepard
interpolation\cite{franke1980smooth,nguyen1995dual,bettens1999learning},
the moving least squares
method\cite{lancaster1981surfaces,ischtwan1994molecular,dawes2008interpolating},
permutation invariant
polynomials\cite{cassam2008symmetry,braams2009permutationally,paukku2013global}
or neural network
approaches\cite{sumpter1992potential,bowman2010ab,jiang2016potential}
to obtain multi-dimensional reactive
PESs.\cite{jordan1995utility,jordan1995convergence,skokov1998accurate,collins2002molecular,duchovic2002potlib,zhang2006ab,li2012communication}
Another approach is based on reproducing kernel Hilbert spaces (RKHS)
which attempts to {\it exactly represent} the energies instead of
finding an acceptable approximation to
them.\cite{ho1996general,hollebeek1997fast,hollebeek1999constructing,MM.rkhs:2017}\\

\noindent
Machine-learning (ML) methods provide estimates for a function value
given input $\mathbf{x}$ (e.g. all Cartesian coordinates of a system)
using a model that was ``trained" on a set of known
data.\cite{rupp2015machine} For intermolecular interactions, the use
of reproducing kernel Hilbert space (RKHS)
theory\cite{aronszajn1950theory} provides means to construct a PES
from a training set based on \textit{ab initio} reference
data.\cite{ho1996general,hollebeek1997fast,hollebeek1999constructing}
Such an approach is typically referred to as kernel ridge regression
(KRR).\cite{hofmann2008kernel,rupp2015machine} The RKHS method has
been successfully applied e.g.\ for constructing PESs for
CNO\cite{MM.cno:2018}, N$_2^+$--Ar\cite{unke2016collision} or
H$_2$O.\cite{ho1996global} A combination of expanding the PES in
spherical harmonics for the angular coordinates and reproducing
kernels for the radial coordinates has been explored for
H$_2^+$--He\cite{MM.heh2:1999} and is now also used for larger
systems.\cite{avoird:2005,avoird:2006}\\

\noindent
To further automatize this process, dedicated computer code has been
made available that generates the interpolation (and meaningful
extrapolation) of the PES along with all required parameters
automatically from girded {\it ab initio} data.\cite{MM.rkhs:2017}
The theory of reproducing kernel Hilbert spaces asserts that for given
values $f_i = f(\mathbf{x}_i)$ of a function $f(\mathbf{x})$ for $N$
training points $\mathbf{x}_i$, $f(\mathbf{x})$ can always be
approximated as a linear combination of kernel products
\cite{scholkopf2001generalized}
\begin{equation}
\widetilde{f}(\mathbf{x}) = \sum_{i = 1}^{N} c_i K(\mathbf{x},\mathbf{x}_i)
\label{eq:RKHS_function}
\end{equation}
Here, the $c_i$ are coefficients and $K(\mathbf{x},\mathbf{x'})$ is
the reproducing kernel of the RKHS. The coefficients $c_i$ satisfy the
linear relation
 \begin{equation}
 f_j = \sum_{i = 1}^{N} c_i K_{ij}
 \label{eq:RKHS_coefficient_relation}
 \end{equation}
with the symmetric, positive-definite kernel matrix $K_{ij} =
K(\mathbf{x}_i,\mathbf{x}_j)$ and can therefore be calculated from the
known values $f_i$ in the training set by solving
Eq.~\ref{eq:RKHS_coefficient_relation} for the unknowns $c_i$ using,
e.g.\ Cholesky decomposition.\cite{golub2012matrix} With the
coefficients $c_i$ determined, the function value at an arbitrary
position $\mathbf{x}$ can be calculated using
Eq.~\ref{eq:RKHS_function}. Derivatives of $\widetilde{f}(\mathbf{x})$
of any order can be calculated analytically by replacing the kernel
function $K(\mathbf{x},\mathbf{x'})$ in Eq.~\ref{eq:RKHS_function}
with its corresponding derivative.\\

\noindent
The explicit form of the multi-dimensional kernel function
$K(\mathbf{x},\mathbf{x'})$ is chosen depending on the problem to be
solved. In general, it is possible to construct $D$-dimensional
kernels as tensor products of one-dimensional kernels $k(x,x')$
\begin{equation}
K(\mathbf{x},\mathbf{x'}) = \prod_{d=1}^{D} k^{(d)}(x^{(d)},x'^{(d)})
\label{eq:multidimensional_kernel}
\end{equation}
For the kernel functions $k(x,x')$ it is possible to encode physical
knowledge, in particular about their long range behaviour. Explicit
radial kernels include the reciprocal power decay
kernel\cite{ho1996general}
\begin{equation}
k_{n,m}(x,x') = n^2 x_{>}^{-(m+1)}\mathrm{B}(m+1,n)_2\mathrm{F}_1\left(-n+1,m+1;n+m+1;\dfrac{
x_{<}}{x_{>}}\right)
\label{eq:reciprocal_power_kernel}
\end{equation}
 or the exponential decay kernel
\begin{equation}
k_{n}(x,x') = \dfrac{n\cdot n!}{\beta^{2n-1}} \mathrm{e}^{-\beta
  x_{>}}\sum_{k=0}^{n-1}\frac{(2n-2-k)!}{(n-1-k)!k!}\left[\beta(x_{>}-x_{<})\right]^k
\label{eq:exponential_kernel}
\end{equation}
where $x_{>}$ and $x_{<}$ are the larger and smaller of $x$ and $x'$
and the integer $n$ determines the smoothness. In
Eq.~\ref{eq:reciprocal_power_kernel} the parameter $m$ is the
long-range decay of the dominant intermolecular interaction
(e.g.\ $m=5$ for dispersion), $\mathrm{B}(a,b)$ is the beta function
and $_2\mathrm{F}_1(a,b;c;d)$ is the Gauss hypergeometric function.\\

\noindent
One particular challenge in extending these methods to larger systems
(tetra- or penta-atomic systems) is therefore to reduce the number of
reference energies while maintaining an accurate representation of the
global PES. Considerable progress in this regard has been recently
made by using either Gaussian Processes combined with Bayesian
inference\cite{krems:2019} or by optimizing permutationally invariant
polynomials (PIPs).\cite{bowman.nma:2019}\\

\noindent
An alternative approach uses the known long-range form of the
interaction potential, a model (e.g. a Morse curve) for the short
range together the statistical adiabatic channel model to determine
capture rates.\cite{troe:1989,troe:1999} Such an approach is
reminiscent of using empirical forms of the potential energy surfaces
for studying the high resolution spectroscopy of van der Waals
complexes.\cite{hutson:1990} One of the advantages over more recent
fitting approaches of reference electronic structure data is the
possibility to examine the role of specific features of the PES on the
observables. As an example, the influence of potential anisotropy on
the reaction rate\cite{troe:1989} or vibrational relaxation can be
examined in a controlled fashion. On the other hand, such an approach
does not necessarily yield a globally valid PES and depends on the
quality of the experimental data.\\

\noindent
For non-reactive collisions (e.g., Ar+CO\cite{MM.arco:2017}), PESs are
computed only for the reactant channel. However, in order to allow
chemical reactions to be described, bonds need to be broken and
formed. A full-dimensional PES describing all the asymptotes/channels
are thus necessary. This is done by mixing the
PESs\cite{ARMD-NagyMeuwly2014} of all possible channels of reactants
and products using smooth switching functions, parametrized in a
fashion as to best capture the potential well and the barrier crossing
regions.\\

\subsection{Nuclear Dynamics}
With global PESs in place, it is then possible to determine
state-to-state cross sections and rates from which total cross
sections and thermal rates can be computed. This information together
with the vibrational relaxation times are the main ingredients for the
CFD and DSMC simulations mentioned in the Introduction. These
quantities can be determined either from quasiclassical trajectory
(QCT) simulations or from numerical solutions of the nuclear
Schr\"odinger equation. For both approaches suitable reviews
exist.\cite{tru79,konthesis,clary:2003}\\

\noindent
{\bf Quasiclassical Trajectories:} In QCT simulations, Newton's (or
Hamilton's) equations of motion are propagated using a numerical
integration in time. The dynamics is governed by the multidimensional
PESs and the initial conditions for $\mathbf{x}$ and $\mathbf{v}$ (or
$\mathbf{p}$ and $\mathbf{q}$) are generated according to a Monte
Carlo scheme. Typical propagators are the velocity verlet integrator
or Runge-Kutta of different orders. The reactant and product
ro-vibrational states are determined following semiclassical
quantization. Since the ro-vibrational states of the product diatom
are continuous numbers, the states are assigned by rounding to integer
values either from histogram binning (rounding to the nearest integer)
or Gaussian binning which weights each trajectory with a Gaussian
shaped function centered on the integer
values.\cite{bon97:183,bon04:106,kon16:4731}\\

\noindent
The state-to-state reaction cross section at fixed collision energy
$E_{\rm c}$ is $\sigma_{v,j \rightarrow v',j'}(E_{\rm c}) = 2 \pi
\int_{0}^{b_{\rm max}} P_{v,j \rightarrow v',j'}(b;E_c) b db$. Monte
Carlo sampling of this integral yields\cite{tru79}
\begin{equation}
\sigma_{v,j \rightarrow v',j'}(E_{\rm c}) = \pi b^2_{\rm max}
\frac{N_{v',j'}}{N_{\rm tot}},
\end{equation}
where $N_{\rm tot}$ is the total number of trajectories, $N_{v',j'}$
is the number of reactive trajectories for final state $(v',j')$, and
$b_{\rm max}$ is the maximum impact parameter for which a reactive
collision occurs. The thermal rate for temperature $T$ is obtained
from
\begin{equation}
 k(T) = g(T)\sqrt{\frac{8k_{\rm B}T}{\pi\mu}} \pi b^2_{\rm max}
 \frac{N_{r}}{N_{\rm tot}},
 \label{eq8}
\end{equation}
where $g(T)$ is the electronic degeneracy factor, $\mu$ is the reduced
mass of the collision system, $k_{\rm B}$ is the Boltzmann constant,
and, depending on the specific process considered, $N_r$ is the number
of reactive or vibrationally relaxed trajectories. In the rate
coefficient calculations, the initial ro-vibrational states and
relative translational energy ($E_{\rm c}$) of the reactants for the
trajectories are sampled from Boltzmann and Maxwell-Boltzmann
distribution at a given $T$, respectively. Such a treatment neglects
the wave nature of the propagation so it is necessary to validate under
what conditions quantum effects are expected to be significant.  \\

\noindent
{\bf Nonadiabatic Effects:} Because at hypersonic conditions the
energetically accessible PESs may cross, it is also relevant to
consider nonadiabatic effects. For describing such transitions several
trajectory-based methods exist. They include, for
example,\cite{billing:1994} fewest switches surface hopping
(FSSH),\cite{Tully90} the Ehrenfest mean field
approach,\cite{mclachlan:1964} or trajectory surface hopping
(TSH)\cite{sti76:3975} within the Landau-Zener
(LZ)\cite{lan32:46,zen32:696} formalism. For the LZ approach the
transition probability $P_{\rm LZ}^{i\rightarrow j}$ from state $j$ to
$k$ is\cite{bel11:014701,bel14:224108,MM.cno:2018}
\begin{equation}
 P_{\rm LZ}^{i \rightarrow j} = {\rm exp} \left( - \frac{\pi}{2\hbar}
 \sqrt{\frac{\Delta V^a_{ij}(R(t_c))^3}{\frac{d^2}{dt^2}\Delta
     V^a_{ij}(R(t_c))}} \ \right).
\end{equation}
where $\Delta V^a_{ij}(R(t_c))$ is the adiabatic energy difference
between states $i$ and $j$ at configuration $R$ and time $t_c$.  In
practice, trajectories are started from a given initial electronic
state $i$. If there is a crossing between the present electronic state
$i$ and a different state $j \neq i$, $P_{\rm LZ}^{i \rightarrow j}$
is calculated and compared with a random number $\xi \in [0,1]$. For
$P_{\rm LZ}^{i \rightarrow j} \geq \xi$ the trajectory hops from state
$i$ to state $j$. To ensure conservation of total energy and total
angular momentum, momentum corrections along different degrees of
freedom have been employed\cite{mil72:5637}
\begin{equation}
 \bf{p'} = p - \hat{n}\frac{\hat{n}M^{-1}p}{\hat{n}M^{-1}\hat{n}}\left
    [{\rm 1}-\left({\rm 1-2\Delta E}
      \frac{\hat{n}M^{-1}\hat{n}}{(\hat{n}M^{-1}p)^{\rm 2}} \right
      )^{\rm 1/2}\right ],
\end{equation}
where ${\bf p}$ and ${\bf p'}$ are the momenta before and after the
hop and $M$ is the mass matrix.\\

\noindent
{\bf Quantum Dynamics:} In the high-temperature limit classical MD
simulations are expected to provide a realistic description for the
dynamics. However, as $T$ decreases, nuclear effects (including zero
point motion, coherence or tunneling) may become more important. The
nuclear Schr\"odinger equation can either be solved within a
time-dependent (TD) or a time-independent (TI)
formalism.\cite{nyman:2000,clary:2003} For the current problem of
determining state-to-state cross sections, quantum reactive scattering
calculations need to be carried out. For reactive and non-reactive TI
scattering calculations, general programs have been made available,
including MOLSCAT,\cite{molscat} Dynasol\cite{dynasol} or
ABC.\cite{abc} In all cases the state-to-state reaction probabilities
are calculated for a partial wave ($J$) and a particular collision
energy $E_c$ from the scattering matrix. The state-to-state cross
sections $\sigma_{v'j' \leftarrow vj}(E_c)$ are then calculated by
summing the probabilities of all partial waves. In each run
probabilities can be calculated for only one $E_c$. Calculating the
state-to-state cross sections as a function of $E_c$ is thus
computationally prohibitive. However, as the time is not directly
involved in the calculations, observables can be calculated down to
very low collision energies.\\

\noindent
Alternatively, in a time-dependent quantum mechanical (TDQM)
approach\cite{kos88:2087,tdqmtannor,konthesis} an initial wavefunction is
written as superposition of wavefunctions (i.e., wave packet) and the
wave packet (WP) is propagated in time and finally, the flux is
calculated at the product/reactant channels to determine the reaction
probabilities. The most widely used representation is based on
Gaussian coherent states which cover a range of energies. To evaluate
the action of the Hamiltonian on the WP it is advantageous to either
work in momentum space and use Fast Fourier transform techniques or to
use the coordinate representation together with a discrete variable
representation (DVR). For propagating the WP in time the
split-operator\cite{fei82:412},
Chebyshev\cite{tal84:3967,che96:3569,gra98:950,che99:19,man95:7390} or
iterative Lanczos techniques can be used.  As a spatially finite grid
is used to represent the WP, reflection from the grid boundary need to
be suppressed by using either complex absorbing potentials or damping
functions at the grid boundary.\cite{man02:9552,pan05:054304} With
these elements in place, the initial WP can be propagated in time and
space until a given final time for which the WP can be projected onto
a final state. Because the WP contains a continuum of collision
energies extending over a finite range, in each run probabilities can
be calculated for a range of energies depending on the WP for a
particular initial state. One of the limitations of a TD approach is
for systems with deep bound wells which require long propagation times
until final states are reached. Similarly, computing reaction
attributes at very low collision energies are challenging due to long
time propagation.\\

\noindent
Quantum dynamical calculations for C-, N-, and O-containing triatomics
have mainly been carried out to determine reaction probabilities and
cross sections at given collision
energies.\cite{pao02:3655,gam03:7156,miq03:3111,gam06:174303,abr09:14824}
However, to calculate the rates, cross sections are summed over a
range of $E_c$ for a given temperature and a large number of partial
waves are needed to converge the cross sections for high collision
energies. Hence, the $J$-interpolated probabilities are used to
calculate the cross sections and rates
thereafter.\cite{miq03:3111,kon16:034303} At high
  temperatures more ro-vibrational states are populated and
 the  translational energy range to cover also increases which adds
  a large number of separate calculations for each
  rotational state ($j$), its $k-$component  
  (projection of $j$ on $z$-axis) and partial waves ($J$) to obtain converged results. Hence, the computation of reaction cross sections
  becomes prohibitive and simulations for hypersonic flow based on data from
quantum dynamics simulations thus is probably intractable.\\

\section{State-to-State, Thermal Rates and Vibrational Relaxation Rates}
As already indicated in the Introduction, hypersonics requires thermal
and state-to-state nonreactive and reactive cross sections and
vibrational relaxation times. These ingredients are then used in more
coarse-grained simulations of the state distribution of the species
involved to characterize the nonequilibrium chemistry around an object
traveling at high speed through a gaseous atmosphere.\\

\noindent
In the following, previous and more recent efforts to determine such
quantities from atomistic simulations and quantum treatments of the
nuclear dynamics are described, in particular for the [NOO],
[NNO], [CNO] systems. A set of particularly relevant
reactions, the Zeldovich or ``thermal NO
mechanism''\cite{zeldovich:1946}, includes the (NO + O / O$_{2}$ + N)
and (NO + N / N$_2$ + O) reactions\cite{Bose1997,Dodd1999} that
describe the oxidation of nitrogen. These reactions, together with a
range of other atom plus diatom and diatom plus diatom reactions form
the core of the 5- and 11-species model used in
hypersonics.\cite{gupta:1990}\\

\noindent
At high temperatures ($\sim 20000$ K), as present in thin regions of
shock layers created at hypersonic speed flight\cite{park:1993}, the
reactive chemical processes can become very complex. Part of the
complexity arises due to higher electronically excited states that
become accessible, and another part is due to thermal
non-equilibrium. When higher electronically excited states need to be
included, 3-dimensional PESs for them need to be calculated, too, and
nonadiabatic transitions between them may become possible.\\

\noindent
{\bf The [NOO] System:} The N($^{4}$S) + O$_2$(X$^{3}\Sigma_g^-$)
$\leftrightarrow$ O($^{3}$P) + NO(X$^{2}\Pi$) reactions are among the
N- and O- involving reactions that dominate the energetics of the
reactive air flow around spacecraft during hypersonic atmospheric
reentry.  To study the dynamics of the forward and backward reactions,
PESs for $^2$A$'$, $^2$A$''$ and $^4$A$'$ electronic states are
necessary.  To compute the thermal reaction rates for the forward and
the reverse reactions, equilibrium constants, PESs for $^2$A$'$ and
$^4$A$'$ electronic states of NO$_2$ are needed.  However, the
$^4$A$'$ state has a barrier of $\sim$1.6 eV for the O+NO collisions
and this state contributes less to the vibrational relaxation. The
$^2$A$''$ state has a deep potential well of $\sim$1.6 eV in the O+NO
channel which leads to efficient vibrational relaxation via complex
formation. PESs for those three electronic states have been calculated
at different level of theory in different works by various
groups.\cite{Says2002,varandas:2003,Mota2012,castro2014computational,MM.no2:2019}\\

\begin{figure}
  \centering
  \includegraphics[scale=0.3]{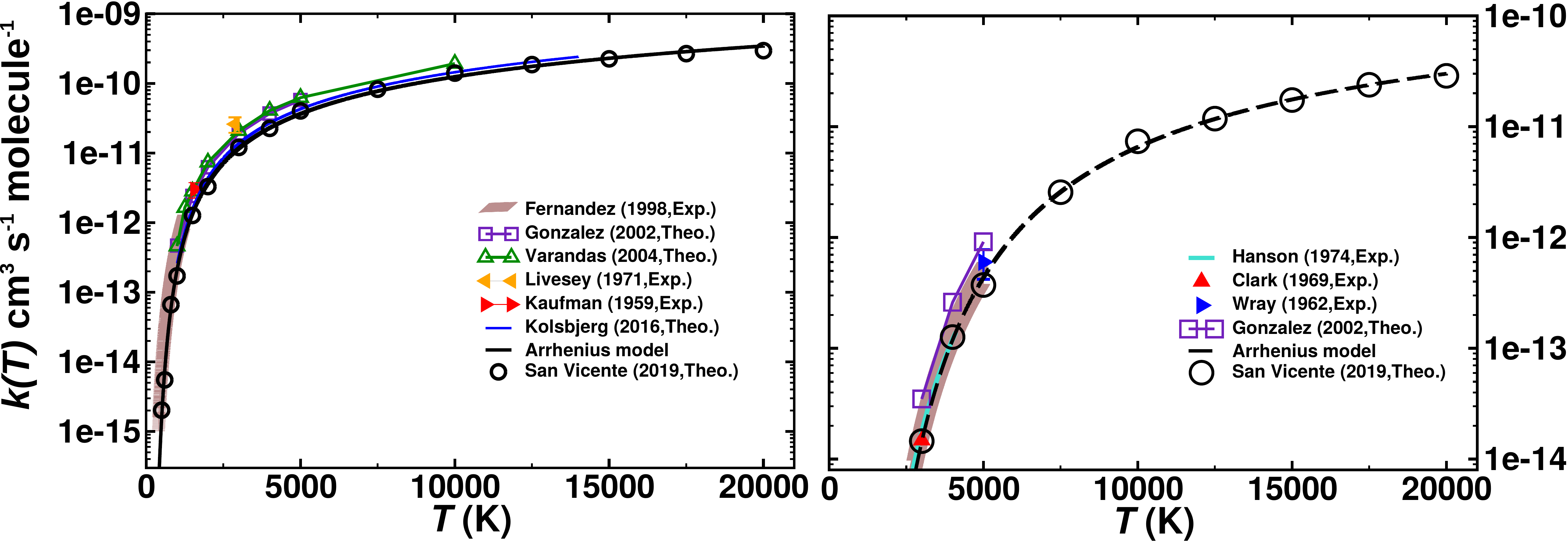}
\caption{Total rates for the N($^{4}$S) + O$_2$(X$^{3}\Sigma_g^-$)
  $\leftrightarrow$ O($^{3}$P) + NO(X$^{2}\Pi$) reactions. The left
  panel shows the rates for the forward reaction and the right panel
  show the rates for the reverse reaction. Arrhenius model are fits of
  the rates from Ref. \cite{MM.no2:2019} to a modified Arrhenius
  model. Data taken from Ref. \cite{MM.no2:2019}.}
\label{fig:rateno2}
\end{figure}

\noindent
Thermal rates have been determined for the forward reaction at many
instances using experiments and
computations\cite{Fernandez:1998,Says2002,Caridade2004,Livesey:1971,Kaufman:1959,hammer:2016,CastroPalacio2015,MM.no2:2019}. However,
for the reverse reaction, theoretical rates\cite{Says2002} are
determined indirectly from the equilibrium constant of N($^{4}$S) +
O$_2$(X$^{3}\Sigma_g^-$) $\leftrightarrow$ O($^{3}$P) +
NO(X$^{2}\Pi$). In Ref. \citen{MM.no2:2019}, explicit QCT calculations
were carried out to calculate the rates for forward and reverse
reactions for a wide range of temperatures which are in good agreement
with the experiment results (see Figure \ref{fig:rateno2}). The
equilibrium constants were also calculated for this reactive system
and found to be in good agreement with the CEA\cite{CEA1996} data
base.  Parameters corresponds to a modified Arrhenius equation are
provided for the forward and reverse reaction rates.\\

\begin{figure}
\centering
\includegraphics[scale=0.5]{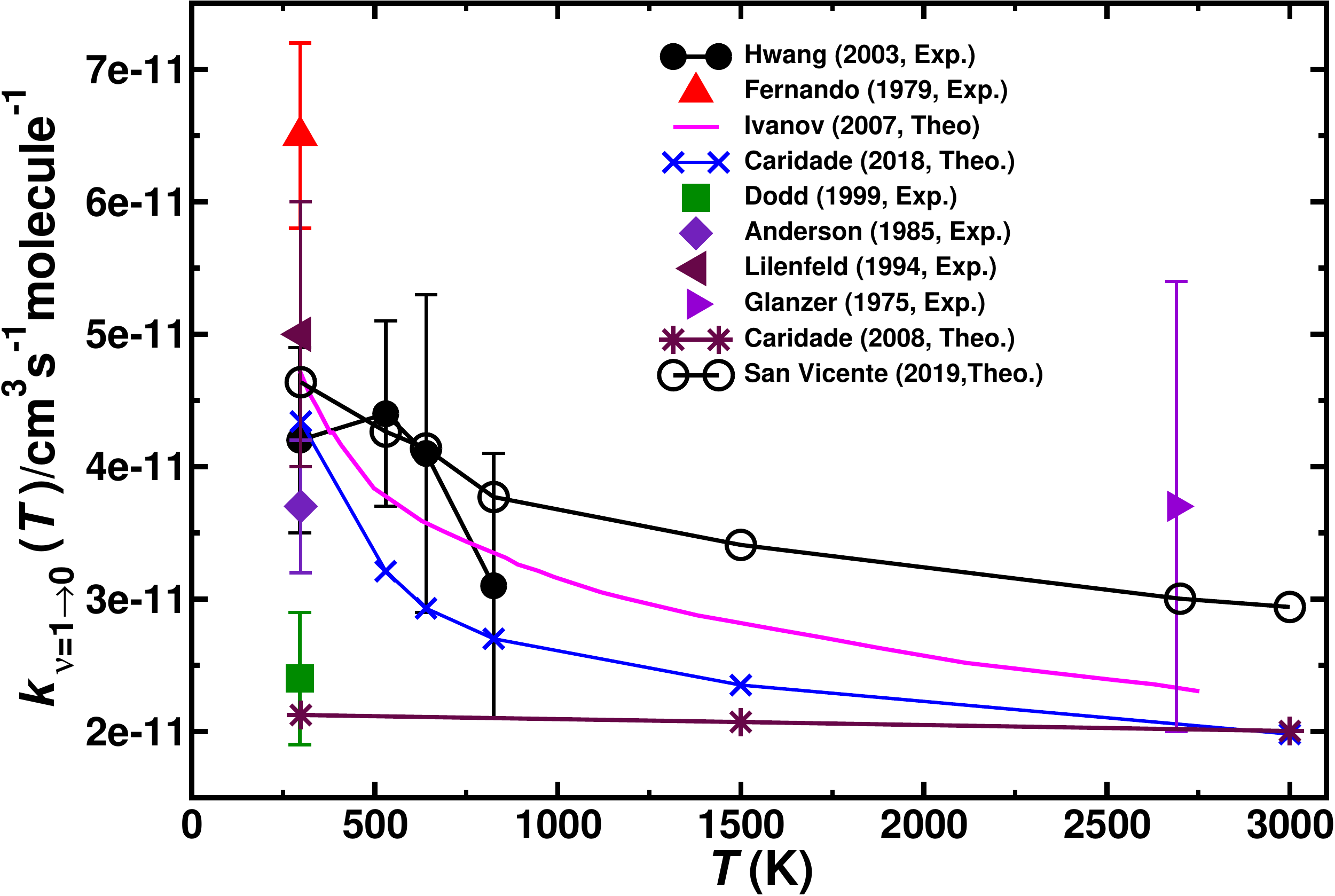}
\caption{Vibrational relaxation rates for O+NO($\nu = 1$)
  $\rightarrow$ O+NO($\nu' = 0$). Symbols with error bars show the
  experimentally determined VR rates and Symbols connected by lines
  show the QCT VR
  rates.\cite{Hwang:2003,Fernando:1979,Schinke:2007,Caridade2018,Dodd1999,Anderson:1985,Lilenfeld:1994,glanzer:1975,Caridade2008,MM.no2:2019}
  Data taken from Ref. \cite{MM.no2:2019}.}
\label{fig:vrno2}
\end{figure}

\noindent
Figure \ref{fig:vrno2} shows the vibrational relaxation rates for the
O+NO collisions for $v=1 \rightarrow v'=0$. Both the nonreactive and
the oxygen exchange collisions contribute to the vibrational
relaxation.  Recent QCT simulations\cite{Caridade2018} based on the
DIM PES for the $^2$A$'$ state\cite{varandas:2003} and a fitted DMBE
PES based on 1681 MRCI/AVQZ calculations for the $^2$A$''$
state\cite{Mota2012} report a value of $k_{\nu=1 \rightarrow 0}(T=298
{\rm K}) = 4.34 \pm 0.7 (10^{-11})$ cm$^3$s$^{-1}$. Another
computational study\cite{Schinke:2007}, using a 3-dimensional spline
representation of energies from CASSCF/aug-cc-pVTZ calculations,
reported a value of $k_{\nu = 1 \rightarrow 0} (T = 300 {\rm K})\sim 5
(10^{-11})$ cm$^{3}$s$^{-1}$. These relaxation rates agree quite
favourably with experiment\cite{glanzer:1975,Hwang:2003} but the
$T-$dependence of the simulations using the DMBE PESs is too steep, in
particular for the $^2$A$''$ state which leads to an underestimation
of the vibrational relaxation at higher temperatures. More recent
simulations based on RKHS-represented PESs at the MRCI/aug-cc-pVTZ
level of theory correctly describe both, the low- and high-$T$
vibrational relaxation rates as observed
experimentally.\cite{MM.no2:2019} \\

\noindent
 {\bf The [NNO] System:} The NO + N $\leftrightarrow$ O + N$_2$
 reactions are important in high temperature gas flows. These
 reactions control the concentration of NO in the hypersonic flow
 during reentry. The two lowest triplet PESs ($^3$A$'$ and $^3$A$''$ )
 of N$_2$O correlate with NO(X$^2\Pi$)+N($^4S$) and
 N$_2$(X$^1\Sigma_g^+$)+O($^3P$).  In the absence of spin-orbit
 coupling, the NO(X$^2\Pi$)+N($^4S$) $\leftrightarrow$ O($^3$P) +
 N$_2$(X$^1\Sigma_g^+$) reactions occur entirely in the triplet
 manifold of N$_2$O. Several {\it ab initio} energy based PESs have
 been constructed to study the reaction dynamics of this system by
 fitting to polynomial functions\cite{gamallo2003ab,lin16:024309} or
 representing by reproducing kernel.  \cite{alp17:2392}\\
 
\noindent
Rate coefficients for the forward and the reverse reactions have been
estimated via experiments using different techniques.  In a discharge
flow-resonance fluorescence (DF-RF) and flash photolysis-resonance
fluorescence the rate for the forward reaction was measured to be $3.4
\pm 0.9 \times 10^{-11}$cm$^3$s$^{-1}$molecule$^{-1}$ over the
temperature range 196--400 K.\cite{lee1978absolute} In two different
shock tube studies,\cite{mic92:3228,mic93:6389} the rates were
estimated over temperature ranges 1850--3160 K and 1251--3152 K as
$3.32\times10^{-11}$ and $3.7\times 10^{-11}$
cm$^3$s$^{-1}$molecule$^{-1}$, respectively.  In a continuous
supersonic flow reactor\cite{bergeat2009low} the rates for the forward
reaction were measured at 48--211 K to be $(3.2\pm0.6)\times 10^{-11}$
exp$(25\pm16/T)$cm$^3$s$^{-1}$molecule$^{-1}$.  For the reverse
reaction, in shock tube experiment, the rates were expressed as
$3.055\times10^{-10}$exp$(38370/T)$ at 2384--3850 K
temperatures.\cite{monat1979shock} In another shock tube
experiment\cite{thielen1985resonance} the rates for the reverse
reaction were measured at 2400--4100 K to be $3.0\times 10^{-10}
\exp{(-38300/T)} \pm 40 \%$ cm$^3$s$^{-1}$molecule$^{-1}$.\\

\begin{figure}
\centering \includegraphics[scale=0.31]{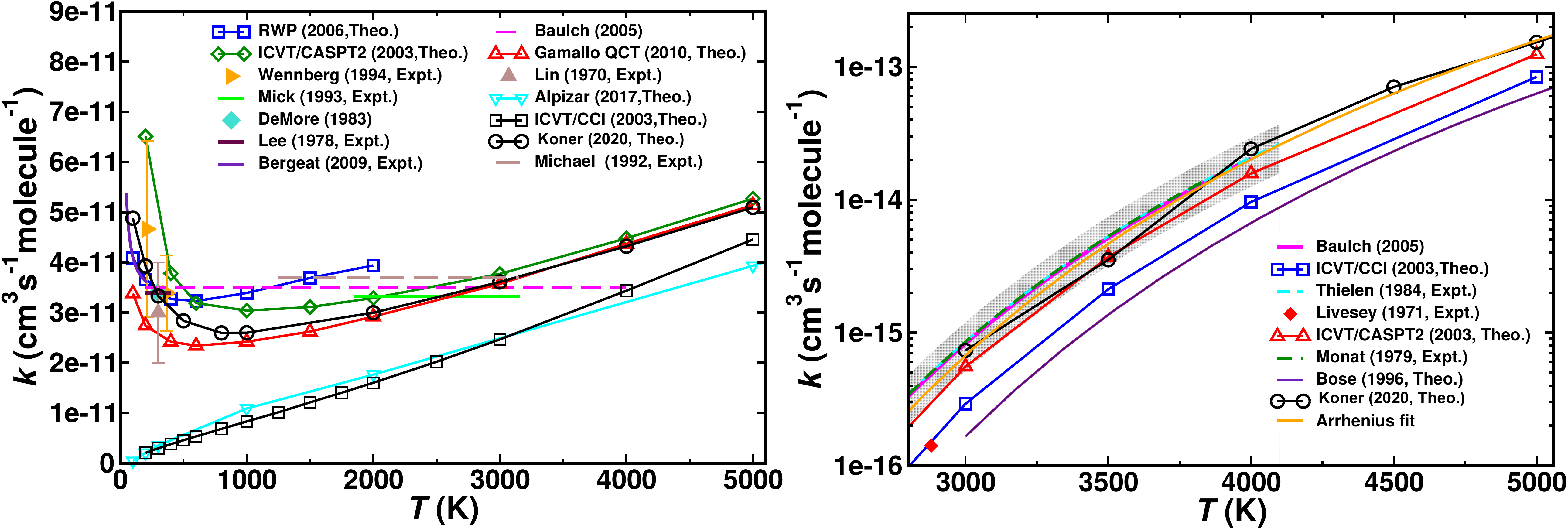}
\caption{Rate coefficients for the NO(X$^2\Pi$)+N($^4S$)
  $\leftrightarrow$ N$_2$(X$^1\Sigma_g^+$)+O($^3P$) as a function of
  temperature. Top panel shows rates for the forward reaction while
  the bottom panel shows the rates for the backward reaction. The
  results calculated in the present work are shown as lines with open
  circles. Experimental (assigned as `Expt.') and Theoretical rates
  (assigned as `Theo.') available in the literature are
  shown.\cite{gamallo2003ab,gam06:174303,gamallo2010quasiclassical,
    alp17:2392,bose1996thermal,lee1978absolute,thi85:685,monat1979shock,
    mic92:3228,mic93:6389,Livesey:1971,bau05,wennberg1994kinetics,bergeat2009low}
  Data taken from Ref. \cite{MM.n2o:2020}.}
\label{fig:n2orates}
\end{figure}

\noindent
Rate coefficients for the forward and the reverse reactions have been
calculated from QCT and quantum simulations for temperatures $100 \leq
T \leq 5000$
K.\cite{gamallo2003ab,gam06:174303,gamallo2010quasiclassical,alp17:2392,bose1996thermal}
Computed rates for both reactions are shown along with the
experimental rates and Baulch recommended
values\cite{lee1978absolute,thi85:685,monat1979shock,mic92:3228,mic93:6389,Livesey:1971,bau05,wennberg1994kinetics,bergeat2009low}
in Figure \ref{fig:n2orates}. For the forward reaction, except the
ICVT/CCI rates from Ref. \citen{gamallo2003ab} and QCT rates from
Ref. \citen{alp17:2392}, good agreement between theory and experiment
is found. At high temperatures rates obtained from all the simulations
are close to each other. The calculations suggest that up to $\sim
5000$ K N$_2$ formation occurs mostly on the $^3$A$''$ PES whereas
reactions involving the $^3$A$'$ state start to contribute at higher
temperatures. High temperature rates up to 20000 K and final state
distributions for the forward reaction have been reported in
Ref. \citen{alp17:2392}. The reverse reaction and the N$_2$
dissociation are also studied recently and rate expressions are
reported.\cite{luo17:074303,esp17:6211}.  In another recent study the
PESs for the two lowest triplet states of N$_2$O have been
reconsidered based on MRCI+Q/aug-cc-pVTZ
calculations.\cite{MM.n2o:2020} The grid was considerably extended, in
particular for the diatomic separation ($r$), and in the long range
interaction region (along $R$) was treated more accurately. For both
reactions, thermal rates are calculated on the new PESs, which are in
good agreement with experiment, see open black circles in Figure
\ref{fig:n2orates}. Upon inspection, the difference between the
previous\cite{alp17:2392} and the improved, more recent
PESs\cite{MM.n2o:2020} is the presence of a small barrier in the N+NO
entrance channel for the $^3$A$''$ state in the former which which
leads to a smaller rates at low temperatures. However, the new
PESs\cite{MM.n2o:2020} show the correct behaviour of $k(T)$ at low
temperatures.\\

\begin{figure}
\centering
\includegraphics[scale=1.2]{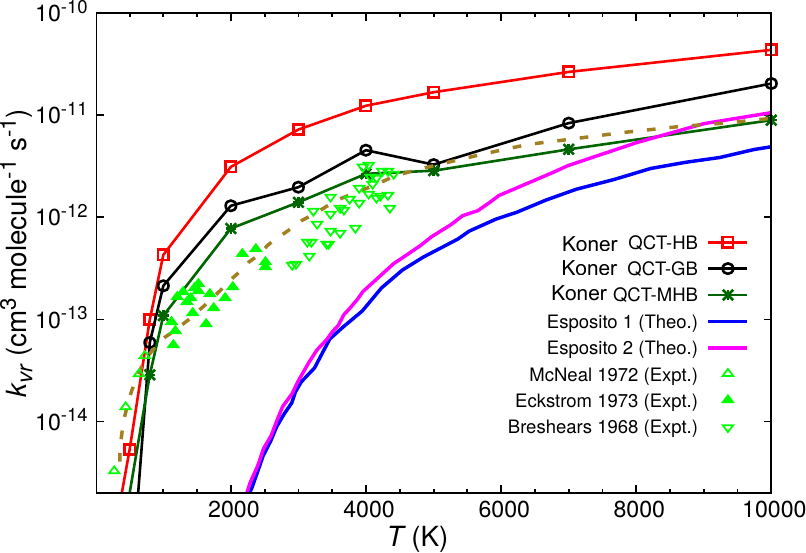}
\caption{Vibrational relaxation rates for O+N$_2$($\nu = 1$)
  $\rightarrow$ O+N$_2$($\nu' = 0$). Green symbols represent the
  experimentally determined VR
  rates.\cite{bre68:4768,mcn72:507,eck73:2787} Olive dashed line is a
  double Arrhenius type fit to the experimental
  result.\cite{on2vrbook} Rates obtained in this work from QCT
  simulations and using HB and GB schemes are given along with the
  full QCT (magenta solid line) and quasi-reactive QCT (blue solid
  line) results from Ref. \cite{esp17:6211}. Data taken from
  Ref. \cite{MM.n2o:2020}.}
\label{fig:vrn2o}
\end{figure}

\noindent
For vibrational relaxation N$_2$($v=1, j$) + O $\rightarrow$
N$_2$($v'=0, j$) + O the rates from the experimental
data\cite{bre68:4768,mcn72:507,eck73:2787} are extracted from the
vibrational relaxation time parameters ($p\tau_{\rm vib}$) following
the Bethe-Teller model\cite{btmodel}. In a recent
study,\cite{esp17:6211} QCT calculated VR rates are found to
significantly underestimate the experimental results. The VR rates
calculated from QCT simulations in this work using the HB and GB
schemes are shown in Figure \ref{fig:vrn2o}. It can be seen in Figure
\ref{fig:vrn2o} that the QCT-GB results are in a fair agreement with
the experiment. However, it is noticed that if the trajectories with
$\varepsilon_{0,j'} \leq \varepsilon_{v',j'} \leq
\varepsilon_{0,j'}+0.075$ eV (0.075 eV $\approx 0.3$ quanta) are
considered as VR trajectories, the results agree well with experiment
(see Figure \ref{fig:vrn2o}, green line with asterisk QCT-MHB). Hence
it is possible that a fraction of the trajectories which do not enter
or remain in the strong coupling region for shorter time exchange only
small amounts of energy and are not fully relaxed.
  However, to further clarify whether this is due to shortcomings in
  the binning strategy, sensitivity analyses of the quality of the
  PESs and quantum simulations are required.\\

\noindent
{\bf The [CNO] System:} Three reactive processes can be considered for
this system: (i) C+NO, (ii) O+CN and (iii) N+CO. The electronic
states, important to study the dynamics of (i)--(iii) collisions are
$^2$A$'$, $^2$A$''$ and $^4$A$''$. The $^2$A$'$ and $^2$A$''$ states
connect the C($^3$P)+NO(X$^2\Pi$), O($^3$P)+CN(X$^2\Sigma^+$) and
N($^2$D)+CO(X$^1\Sigma^+$) channels while the $^4$A$''$ state connects
the C($^3$P)+NO(X$^2\Pi$), O($^3$P)+CN(X$^2\Sigma^+$) and
N($^4$S)+CO(X$^1\Sigma^+$) channels.  The C+NO $\rightarrow$ O+CN,
N+CO reactions play crucial role in removing NO from the atmosphere
(``NO reburning'')\cite{lamoureux:2016} and the CN+O and CO+N
reactions are important for combustion in flames and for entry into
the atmospheres of Mars or Venus.\cite{brandis:2014} \\

\noindent
Thermal rates for the C($^3$P)/C($^1$D) + NO $\rightarrow$ O + CN were
measured experimentally to be $7.3 \pm 2.2 \times 10^{-11}$ cm$^3$
molecule$^{-1}$ s$^{-1}$ in the gas phase at room
temperature\cite{hus71:543} and later recalculated as 4.8 $\pm$ 0.8
$\times$ 10$^{-11}$ cm$^3$ molecule$^{-1}$ s$^{-1}$.\cite{hus75:525}
In a shock tube experiment, the rates and branching ratios of products
were measured for the C+NO reaction over temperature range of 1550
\textendash\ 4050 K.\cite{dea91:3180} and found to be constant. Rate
coefficients for the same reaction were found to decrease with
increasing temperature.\cite{cha00:8466} Analytical PESs have been
constructed form CASPT2 energies for the calculations and the same
parametrization of the PES were constructed for the $^2$A$'$,
$^2$A$''$ and $^4$A$''$ electronic states of CNO
\cite{sim95:141,and00:613,and00:99,and08:4400} and subsequent dynamics
by means of QCT \cite{and00:613,and00:99,and03:5439,and08:4400} and
adiabatic capture calculations\cite{fra12:4705} result thermal rates
close to the experimental ones but the branching ratios of CO and CN
products underestimate the experimental values. A more accurate DMBE
PES for the $^2$A$'$ electronic state of [CNO] has been computed at
MRCI-F12/cc-pVQZ-F12 level of theories and quasiclassical dynamics
have been carried out but not compared with the experiment because a
single PES is not sufficient to describe the
dynamics.\cite{gon18:4198,alv19:7195}\\

\begin{figure}
  \centering \includegraphics[scale=1.05]{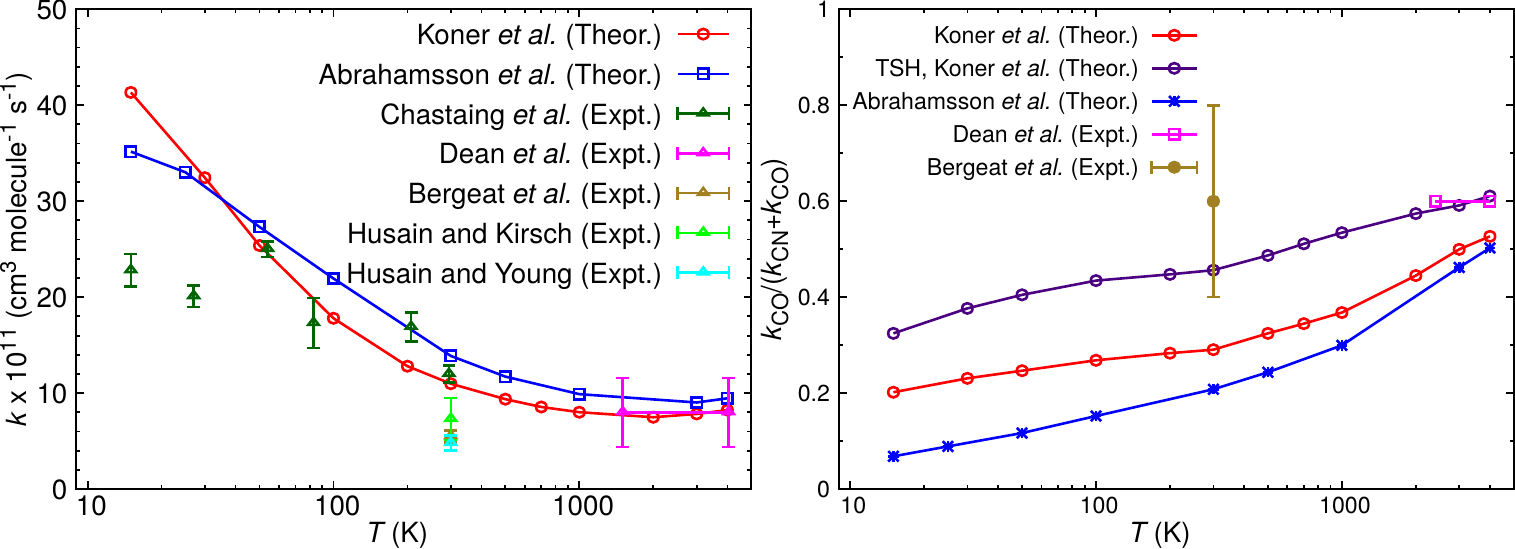}
\caption{Left panel: Total rates for the C($^3$P)+NO(X$^2\Pi$)
  $\rightarrow$ O($^3$P)+CN(X$^2\Sigma^+$),
  N($^4$S/$^2$D)+CO(X$^1\Sigma^+$) reaction. Right panel: Branching
  fraction (CO vs CN) for the same reaction.  TSH represents the
  results from trajectory surface hopping dynamics. `Theor.'
  represents the results obtained from
  computations\cite{and08:4400,MM.cno:2018} while `Expt.' represents
  the experimental observations.\cite{
    hus71:543,hus75:525,dea91:3180,ber99:7,cha00:8466} Data taken from
  Ref. \cite{MM.cno:2018}}
\label{fig:ratecno}
\end{figure}

\noindent
In recent work,\cite{MM.cno:2018} $\sim 50000$ MRCI+Q/aug-cc-pVTZ
energies have been used to construct accurate RKHS-based
representations of the 3 global PESs for the $^2$A$'$, $^2$A$''$ and
$^4$A$''$ states of [CNO] with root mean squared errors of 0.38, 0.48
and 0.47 kcal/mol, respectively. Subsequent quasiclassical dynamics
study on those RKHS PESs yield thermal rates which are plotted in
Figure \ref{fig:ratecno}. The rates agree well with experiment for
temperatures between 50 K and 5000 K.  The branching fraction of the
CO product for the C+NO reaction is also shown in Figure
\ref{fig:ratecno}. Including nonadiabatic transitions improve the
branching ratios to be compared with the experimental findings. The
vibrational relaxation rates for C($^3$P) + NO(X$^2\Pi$)($v=1,j$)
$\rightarrow$ C($^3$P) + NO(X$^2\Pi$)($v'=0,j$) are also
calculated. Explicit fit to Arrhenius equation are provided for the
C+NO reaction up to 20000 K.\\

\begin{table}[h]
\caption{ Parameters obtained by fitting the rates for different
  reactions to a modified Arrhenius equation
  ($k(T)=AT^{n}$exp$(-E_a/T)$).  Rate coefficients computed using
  these parameters have units in cm$^3$molecule$^{-1}$s$^{-1}$ with
  [$A$] = cm$^3$molecule$^{-1}$s$^{-1}$ and [$E_a$] = K while $n$ is
  unitless.}
\begin{small}
\begin{tabular}{l|rcrr}
\hline
\hline
\ \ \ \ \ \ \ \ \ \ \ \ \ \ Reaction &$T$(K) \ \ \ \  &$A$ & $n$ \ \ \ \  & $E_a$  \ \ \  \\
\hline
O($^3$P)+NO(X$^2\Pi$)$\rightarrow$N($^4$S)+O$_2$(X$^3\Sigma_g^-$) & 3000--20000 & $7.95872\times10^{-13}$ &0.48656 & 23749.0 \\
N($^4$S)+O$_2$(X$^3\Sigma_g^-$)$\rightarrow$O($^3$P)+NO(X$^2\Pi$) & 500--20000 & $3.70470\times10^{-15}$ & 1.17593 & 4090.1  \\
\hline
N($^4$S)+NO(X$^2\Pi$)$\rightarrow$O($^3$P)+N$_2$(X$^1\Sigma_g^+$) & 2000--18000 & $2.17214\times10^{-14}$ & 0.88796 & -946.0 \\
O($^3$P)+N$_2$(X$^1\Sigma_g^+$)$\rightarrow$N($^4$S)+NO(X$^2\Pi$) & 3000--20000 & $7.73865\times10^{-12}$ & 0.46177 & 39123.1  \\
O($^3$P)+N$_2$(X$^1\Sigma_g^+$)$\rightarrow$2N($^4$S)+O($^3$P) & 8000--20000 &4.55027 & --2.00227 & 129692.6 \\
\hline
C($^3$P)+NO(X$^2\Pi$)$\rightarrow$O($^3$P)+CN(X$^2\Sigma^+$) & 5000--20000 & $2.96396\times10^{-13}$ &0.55000&-1043.3\\
C($^3$P)+NO(X$^2\Pi$)$\rightarrow$N($^2$D)+CO(X$^1\Sigma^+$) & 2000--20000 & $3.76536\times10^{-13}$ &0.49438&-729.1\\
C($^3$P)+NO(X$^2\Pi$)$\rightarrow$N($^4$S)+CO(X$^1\Sigma^+$) & 700--18000 & $5.93636\times10^{-15}$ &0.94205&-607.7\\
\hline
\hline
\end{tabular}
\end{small}
\label{tab:afit}
\end{table}

\noindent
Table \ref{tab:afit} summarizes the parameters from a modified
Arrhenius equation $k(T)=AT^{n}$exp$(-E_a/T)$ which was fitted to the
rates for different reactions containing C-, N-, and O-species.  The
rates were computed from extensive QCT calculations on accurate RKHS
PESs based on MRCI+Q/aug-cc-pVTZ {\it ab initio}
energies.\cite{MM.cno:2018,MM.no2:2019,MM.n2o:2020} For all the
reactions the rates are in good agreement with available experimental
results (see Figures \ref{fig:rateno2}, \ref{fig:n2orates}, and
\ref{fig:ratecno}) over the temperature range $\sim 50-5000$ K and
allow predictions at considerably higher ($\sim 20000$ K).\\

\section{Outlook}
Up to this point the necessary microscopic dynamics from which cross
sections, thermal and vibrational relaxation rates was determined ``on
demand'' from a given set of initial states by running explicit QCT
simulations. However, such an approach can be computationally
prohibitive in multi scale simulations, such as DSMC which attempt to
solve a spatio-temporal chemical model by decomposing space around an
object into discrete cells of different dimensions (``voxels''). In
each of the voxels chemical processes can occur and the necessary
information for modeling the temporal and spatial evolution needs to
be determined from either explicit QCT simulations or from evaluating
a simplified model.\\

\noindent
As the number of particles ranges from $10^5$ to $10^{10}$ and the
simulation time scales are macroscopic, efficient models are
required. Under such conditions, running direct QCT simulations
becomes unfeasible as there are $\sim 10^4$ internal $(v,j)$ states
for a diatomic molecule which leads to $\sim 10^{15}$ state-to-state
transitions for diatom-diatom collisions.\cite{grover:2019} One
possibility consists of developing more coarse-grained models either
by averaging over rotational energies, or by using energy-binning
strategies,\cite{panesi:2018} to reduce the number of simulated
transitions. However, it has been found that depending on the way how
this coarse-graining is carried out, the internal energy
distributions, relaxations and dissociation rates can be markedly
different.\cite{andrienko:2016,torres:2018} As an alternative, the
direct molecular simulation (DMS) method has been
developed.\cite{koura:1997,valentini:2018}\\

\noindent
One recently explored possibility is to train a machine learned model
based on neural networks from explicit QCT data for state-to-state
cross sections from which all necessary information can be
determined.\cite{MM.nn:2019} Such an approach combines the accuracy of
QCT simulations based on state-of-the art electronic structure
calculations and PES representation techniques with the necessary
speed to obtain the molecular-level data for nonreactive and reactive
atom+diatom collisions. For this, the N($^4$S)+NO(X$^2\Pi$)($v,j$)
$\rightarrow$ O($^3$P)+N$_2$(X$^1\Sigma_g^+$)($v',j'$) reaction has
been considered to model the state-to-cross sections on $^3A'$
PES. There are 6329 ro-vibrational states for the N+NO channel, and
8733 states for the O+N$_2$ channel giving rise to $\sim 10^7$
state-to-state transitions.\\

\begin{figure}
\includegraphics[scale=1.35]{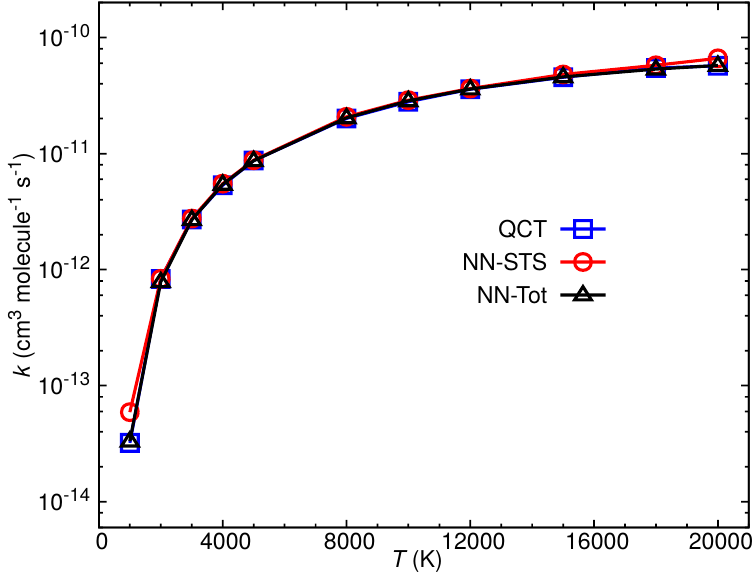}
\caption{Total rates for the N + NO $\rightarrow$ O + N$_2$ reaction
  calculated from QCT simulations on the $^3A'$ state (blue) and
  predicted by the NN models (NN-state-to-state - red and NN-total -
  black). Data taken from Ref. \cite{MM.nn:2019}}
\label{nnrates}
\end{figure}

\begin{figure}
\includegraphics[scale=1.25]{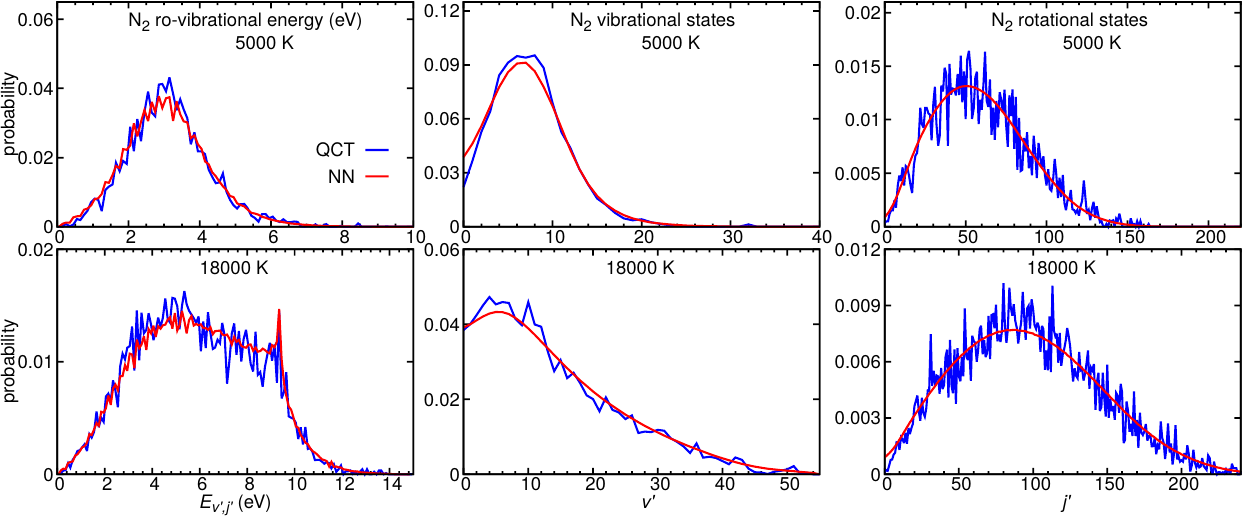}
\caption{Distributions of product ro-vibrational states and
  ro-vibrational energies at 5000 and 18000 K for the N + NO
  $\rightarrow$ O + N$_2$ reaction calculated from QCT on the $^3A'$
  PES (blue) and predicted by the NN state-to-state model (red),
  respectively. Data taken from Ref. \cite{MM.nn:2019}}
\label{nndist}
\end{figure}

\noindent
Using exhaustive QCT simulations on a subset of the total state space
a NN was trained based on the ResNet
architecture.\cite{he16:770,MM.physnet:2019} Importance sampling of
impact parameter in the QCT simulations and averaging the neighbour
states contributions to the cross sections drastically reduces the
number of trajectories required for converge results. Two different NN
model were constructed (i) based on state-to-state cross sections and
(ii) based on total cross sections. The validity of the NN was
established by predicting observables obtained from the NN and
explicitly calculating them using QCT simulations. The NN model
successfully captures the trends as well as the magnitudes of the
rates from QCT. For most cases the relative errors are $< 5$\%
although for individual states they can differ by up to 17 \%. Total
rates $k(T)$ calculated from QCT simulations and predicted by the NN
models are shown in Figure \ref{nnrates}.  In this case the agreement
is within a few percent. As another test, the distribution of the
final vibrational and rotational states and the rovibrational energies
of N$_2$ after N+NO collisions at different temperatures was
calculated from QCT and compared with those from the NN. The NN
correctly captures the shape of all distributions but lacks
oscillatory features, in particular for the rotational distribution
(see Figure \ref{nndist}).  It was hence found that the NN provides a
physically robust model based on validated, microscopic data from
which information about nonequilibrium systems can be obtained,
obviating the construction of models based on simple, empirical
expressions.\cite{schwartz:2018} As the evaluation time of the NN is
on the order of seconds for $10^6$ state-to-state cross sections, this
technique is suitable for direct use in DSMC simulations . The average
error from the NN compared with the reference QCT data is $\sim 5$
\%. This compares with errors ranging from 25 \% to 60 \% for
vibrational relaxation rates and state-specific dissociation rates
from a maximum entropy model for O$_2$ + O.\cite{alexeenko:2016}\\

\noindent
Computing and learning the state-to-state cross sections for
bi-molecular collisions is a tedious task as the number of possible
transitions increases rapidly. For diatom-diatom collision systems the
problem becomes intractable.\cite{valentini:2018} One possibility to
avoid this is to train distributional models based on initial and
final states and energy distributions at different ro-vibrational and
translational temperatures using machine learning, similar to the
model for state-to-state cross sections discussed above.\\

\noindent
Hence from combining expertise and computational strategies rooted in
different disciplines across chemical physics and computational
chemistry it is expected that realistic, robust and computational
tractable models based on accurate molecular processes can be built
for reactive, rarefied flows at different thermodynamic conditions,
including the hypersonic regime. Such a model still requires
approximations to be made. However, by using the highest possible
level of theory at each step it is also expected that meaningful and
informative error estimates can be provided as to the reliability of
the models. One example is the question how sensitive the results of
the QCT (or also quantum dynamics) simulations are to the local and
global shape and quality of the PESs. Such sensitivity analyses can be
computationally demanding in itself but become possible with the
increased computational resources available.\\

\noindent
As the field of reactive A+BC collisions continues to mature,
quantitative assessment of the reliability and predictability of the
underlying PESs and the type (quantum vs. classical) of dynamics
become important in particular if the results are used in reaction
networks or more coarse grained simulations. Every element in the
chain from electronic structure calculations, coverage of
conformational space, representation/fitting of the points,
QCT/quantum simulations, and determining cross sections/rates from
them has its own errors associated with it. Hence, when using rates or
cross sections as input to more coarse grained treatments of reaction
networks it is highly desirable to have realistic error estimates of
the individual steps. This also provides the basis for targeted
improvements of the data and input on which the more coarse grained
simulations are based.\\

\noindent
If sufficient high-quality experimental data is available, one
promising tool that has been tested for high-resolution spectroscopy
is the ``morphing potential'' approach.\cite{MMmorphing99} It directly
relates the PES with the observables and obviates all intermediate
steps. However, to the best of our knowledge, this has never been
attempted for reaction or vibrational relaxation rates.\\

\noindent
In conclusion, describing reaction and vibrational relaxation rates
and state-to-state cross sections relevant to conditions is a
formidable problem spanning several length and temporal scales. For
meaningful calculations and input data useful to more coarse grained
simulations the best methods affordable at every step are
required. With such tools in hand, progress can be made in this
challenging and multifaceted field of physico-chemical relevance.\\

\begin{acknowledgement}
This work was supported by the Swiss National Science Foundation
grants 200021-117810, 200020-188724, the NCCR MUST, and the University
of Basel. \\
\end{acknowledgement}

\providecommand{\latin}[1]{#1}
\providecommand*\mcitethebibliography{\thebibliography}
\csname @ifundefined\endcsname{endmcitethebibliography}
  {\let\endmcitethebibliography\endthebibliography}{}

\end{document}